\documentclass[acmsmall]{acmart}
\usepackage{longtable}
\usepackage{colortbl}
\usepackage{multirow}
\usepackage{subcaption}
\usepackage{csquotes}
\usepackage{color}
\usepackage{tabularray}

\NewDocumentCommand{\pquote}{+O{} +m}{\blockquote[#1]{\textit{#2}}}

\newcommand{\elide}[1]{\textelp{}} % produces ... instead of argument
 
\newcommand{\textcite}[1]{\citeauthor{#1}~\cite{#1}}

\begin{document}

\title{Understanding and Facilitating Mental Health Help-Seeking of Young Adults: A Socio-technical Ecosystem Framework}

\author{Jiaying Liu}
\affiliation{%
  \institution{The University of Texas at Austin}
  \country{USA}}
\email{jiayingliu@utexas.edu}

\author{Yan Zhang}
\affiliation{%
  \institution{The University of Texas at Austin}
  \city{Austin}
  \country{USA}
}
\email{yanz@utexas.edu}
\renewcommand{\shortauthors}{}

\begin{abstract}
Prior research on young adults' mental health help-seeking mostly focuses on one particular resource such as a mobile app or digital platform, paying less attention to their lived experiences interacting with the ecosystem of resources. We conducted in-depth interviews with 18 participants about their help-seeking and non-help-seeking experiences. Guided by Social Ecological Theory, we proposed a Socio-technical Ecosystem Framework for mental health care, consisting of four levels of resources, including technological-, interpersonal-, community-, and societal level resources. Using this framework, we identified two types of support systems for help-seeking, single-resource support system and multi-resource support system. These resources support young adults' help-seeking via three mechanisms, \textit{care-giving}, \textit{care-mediating}, and \textit{care-outreaching}, forming various pathways to care. We then pointed out the barriers to resource use at each level and the general challenges in finding a support system. Our findings contributed to a conceptual framework to categorize mental health care. It also serves as a practical framework to identify challenges in the pathways to care and discover design implications.
\end{abstract}



\maketitle
\section{Introduction}
The Mental Health crisis in the United States is on the rise, especially after the Covid-19 pandemic. 
According to the National Institute of Mental Health (2023) \cite{national_institute_of_mental_health_mental_2023}, around 57.8 million U.S. adults live with a mental illness in 2021, taking up 22.8\% of all U.S. adults. Young adults between 18-25 have the highest prevalence of mental health concerns (33.7\%) compared with other age groups (e.g., 26-49 (28.1\%) and >50 (15\%)), but have the lowest rate of receiving outpatient mental health treatments. Exploring effective strategies for promoting help-seeking behaviors is essential to enhance the well-being of young adults and other populations \cite{hunt_mental_2010}.

%Our study focuses on young adults aged 18-25, as they exhibit the highest prevalence of mental health concerns (33.7\%) compared to other age groups, yet they have the lowest outpatient mental health treatment utilization rates \cite{national_institute_of_mental_health_mental_2023}. Young adults, at a critical developmental stage, face amplified stress and anxiety, due to significant life tasks, including striving for independence, identity formation, and career establishment; however, their pursuit of assistance is obstructed by limited resources, alongside financial, geographic, and transportation barriers to essential mental health services.

A significant portion of existing research in HCI and CSCW related to mental health adopts a technological perspective, focusing on designing mobile apps \cite{wiljer_effects_2020, alqahtani_co-designing_2021}, digital platforms \cite{wong_postsecondary_2021}, and wearable sensors \cite{hickey_smart_2021} to facilitate self-management \cite{petelka_being_2020} and online group interactions \cite{zhang_online_2018}. However, the real-world adoption rate of these tools remains relatively low. For example, a study surveyed 662 students from two U.S. Midwestern colleges, indicating that only 10.1–13.8\% had experience with such digital technologies \cite{toscos_college_2018}. The engagement with such online resources is also limited, primarily involving online searches and information seeking \cite{pretorius_searching_2020}. Some scholars have suggested that the disconnection of these tools from the everyday lives of young adults may be a potential reason for their relatively low ecological validity \cite{mohr_three_2017}.

In the actual practice of seeking help, young adults navigate a diverse ecosystem of resources spanning formal treatment to informal social support, and online sources to offline relationships \cite{burgess_technology_2021}.  A variety of resources play distinct roles in different stages and contexts of mental health help-seeking \cite{cornish_meeting_2017}. 
Yet, an in-depth comprehension of the ecosystem of mental health resources is crucial to identifying opportunities and barriers for developing valid behavior change interventions \cite{johnson_technology-based_2022}. 
This understanding not only ensures mental health technology adoption but also maximizes the financial and social benefits of investing in these technologies. 
%Therefore, this study aims to gain a holistic understanding of how young adults interact with the ecosystem of mental health resources in the help-seeking process.

%
%For instance, \textcite{bhattacharjee_investigating_2023} emphasized the importance of selecting the right location and time of sending text messages to improve users' wellbeing. 

We applied the Social Ecological Theory \cite{stokols_translating_1996} as our guiding framework to explore young adults' lived experiences interacting with the broad and dynamic ecosystem of resources during mental health help-seeking. We specifically focused on their resource use practices since it is the essential and most central part of mental health help-seeking \cite{toscos_college_2018}. The research questions are:
\begin{itemize}
    \item RQ1: What are young adults' resource use practices when seeking help for their mental health concerns? 
    %Specifically, a) What sources do young adults use, and b) What are the relationships among different kinds of resources?
    %\item RQ2: What affordances of resources influence their choice in help-seeking?
    \item RQ2: What are the barriers to young adults' mental health help-seeking?
\end{itemize}

%\cite{ralston_fulfilling_2019} calls for expanding health technology and collecting feedback from underserved communities. built a RE-AIM model synthesizing the current objectives and efforts put into technology-mediated mental health, including Reach, Effectiveness, Adoption, Implementation, and Maintenance. 

A \textbf{Socio-Technical Ecosystem Framework} emerged from this study. We iteratively interviewed 18 young adults with varying mental health concerns about their help-seeking practices. In the data analysis, we focused on resource use in young adults' mental health help-seeking process under the guidance of the Social Ecological Theory. 
Our proposed Socio-technical Ecosystem Framework of mental health resources presents the hierarchical resources consisting of four levels (i.e., societal, community, interpersonal, and technological) that cover professional service, informal help, digital mental health tools, etc. 

Using this framework, we further identified two types of support systems of help-seeking, single-resource support system, and multi-resource support system, which helps evaluate the stability of the support system of an individual. These resources support help-seeking via three mechanisms, \textit{care-giving}, \textit{care-mediating}, and \textit{care-outreaching}. We illustrated how the three mechanisms function in the lived experiences of participants by analyzing their pathways to care in the ecosystem. We then summarized the barriers at each level and general challenges in young adults' practices to find a support system. 
%This practical framework to analyze young adults' mental health help-seeking process, identify potential barriers, and enlighten future redesign of the purposes, roles, and functions of digital technologies, but structures of and connections among levels of resources.
Our contributions include:
\begin{itemize}
    \item We developed a Socio-technical Ecosystem Framework for mental health help-seeking resources, providing a conceptual understanding of the hierarchy of four levels of resources and three mechanisms of support.
    \item We demonstrated how to leverage this unifying framework to categorize individuals' support systems, analyze their pathways to care, and identify barriers during help-seeking.
    \item We presented actionable design implications for mental health technologies, and outlined further directions to engage various levels of resources.
\end{itemize}

%\textbf{Terminology}. According to \textcite{keyes_subjective_2006}, we use \pquote{mental illness} to cover a \pquote{full range of diagnosable mental illnesses and disorders,} and \pquote{mental distress} to refer to experiences and symptoms of mental illness without a psychiatric diagnosis. 

\section{Related Work}
We reviewed the literature concerning the models and practices of young adults' mental health help-seeking, especially around resource use.

\subsection{Mental health help-seeking process and behavioral models} 
Mental health help-seeking refers to people's behavior of actively obtaining assistance from various sources to cope with problems and experiences triggered by mental distress. The assistance includes suggestions, information, and treatment that serve pragmatic purposes and also social and emotional support and understanding \cite{lachmar_mydepressionlookslike_2017}. 

One of the earliest models focused on formal help-seeking in the medical system. \textcite{huxley_mental_1996} outlined pathways that a patient may need to navigate through to get care. The pathway consists of five layers. Potential patients entered the pathway from general practitioners, via whom they were referred to primary care, then secondary psychiatric care, and ended with admission to the hospital. The following studies identified other resources that act as referrals within \cite{gater_pathways_1991} and outside \cite{bhui_mental_2002} of the medical system, such as native or religious healers, nurses, psychiatric services, and police institutions.
The model does not sufficiently consider external factors such as socio-economic status, cultural influences, or the role of informal support networks, which can significantly impact the care-seeking journey. 

Rickwood’s Stages of help-seeking framework \cite{rickwood_young_2005} describes patients' subjective experiences of the help-seeking process. The model delineates four key steps: (1) recognizing symptoms and realizing the need for assistance, (2) expressing symptoms and signaling the need for support, (3) identifying available and accessible sources of help, and (4) the final step contingent on the individual's willingness to disclose difficulties to the chosen source. The stages are presented as static categories, potentially overlooking the dynamic nature of help-seeking over time. 

Seeking help for mental health is not a quick and simple decision; rather, it is a complex and dynamic process dubbed with stigmatization and solitary feelings \cite{lannin_does_2016}. In the process, people constantly negotiate with themselves "if seeking help is necessary" and "when and how to seek help". \textcite{biddle_explaining_2007} delineated this struggle through the Cycle of Avoidance (COA) model, depicting the tensions involved in making sense of, accepting, and avoiding mental distress. They identified several key actions that postpone help-seeking, including normalizing symptoms, proposing alternative explanations, accommodating increasingly severe distress, pushing the threshold between “normal” and “real” distress, and delaying help-seeking. Other studies also pinpointed similar cognitive strategies such as utilizing self-resilience, denying the effects of professional help, and problematizing help-seeking \cite{martinez-hernaez_non-professional-help-seeking_2014, abavi_exploration_2020}. 

While these behavioral models adeptly capture the intricacies of mental health help-seeking dynamics, they remain somewhat abstract and descriptive. To provide concrete implications that could enhance the effectiveness of mental health help-seeking, this study aimed to focus on resource utilization in help-seeking practices.

%These empirical studies and theoretical models mainly come from the medical and psychiatry fields, and thus focus on characterizing the behaviors of and barriers to young adults' help-seeking. Thus, this study tries to integrate the psychological model into the technology used in mental health help-seeking. 

\subsection{Resources use of young adults' mental health help-seeking}
Studies have pointed out that resource use is a central practice that supports various needs in help-seeking \cite{toscos_college_2018}. One of the most used classifications of mental health resources comes from \textcite{rickwood_conceptual_2012}'s model, identifying four types of help: formal, semi-formal, informal, and self-help. Formal help-seeking involves structured professional assistance within established healthcare systems including professionals, medical treatment, and public institutions. Semi-formal avenues engage sources that offer some structure but may not be professionally trained, like teachers or school counselors. Informal help-seeking relies on personal relationships, involving friends and family. Informal sources play a significant role in mental health help-seeking, serving as crucial channels through which individuals seek assistance and support for their mental well-being \cite{amato_exploratory_1985}. Family and friends, in particular, often serve as the first point of contact for those facing mental health challenges \cite{rickwood_young_2005}. Lastly, self-help signifies individual efforts, from reading self-help books to utilizing online resources. 
However, the four categories are not well defined in this model but are exemplified with instances. Moreover, the growing mental health technologies are not well presented in this framework.

%The landscape of mental health resources has both expanded and diversified with increasing public engagement and policy attention.
The HCI and CSCW community endeavor to innovate various technologies to support clinic treatment and self-care for mental health \cite{ralston_fulfilling_2019}. 
%Technologies such as social media platforms, video-sharing platforms, and mobile applications, also contribute to intricate and evolving pathways to care. 
%A systematic review \cite{johnson_technology-based_2022} found the majority of studies experimented on Internet-based apps and multi-component intervention is needed.
Our review mainly identified four types of mental health technologies:
(1) Search engines and websites that aggregate relevant information about mental health;
%increase people's intentions to seek formal help-seeking \cite{drydakis_m-health_2022};
%Results from an Australian national survey have suggested that accessing information on the Internet is associated with increased use of mental health services and professionals \cite{reavley_sources_2011}. 
%The second mechanism is 
(2) Self-management tools that facilitate self-learning and self-tracking, allowing individuals to document and contemplate their health behaviors \cite{wiljer_effects_2020};
%The third mechanism is 
(3) Online communities that foster peer interactions and enhance social support;
(4) Communication and decision-supporting technologies in clinical settings that facilitate patient education \cite{rauschenberg_compassion-focused_2021}, video consultations \cite{tonnies_health_2021}, and diagnosis and treatment. 

Andersen's model  \cite{andersen_behavioral_1968,andersen_revisiting_1995,babitsch_re-revisiting_2012} further identified three clusters of factors that affect health resource utilization. The first cluster is predisposing factors, such as sociodemographic variables (e.g., age and gender), and health-related beliefs and knowledge. These factors shape individuals' attitudes and perceptions toward seeking healthcare. The second cluster is enabling factors, encompassing logistical elements that facilitate or hinder access to care, such as financial resources and social support. The cluster, need-related factors, including symptoms of psychological distress, drives the actual demand for healthcare services. 
The following studies had slightly different categories of factors, for example, \textcite{lu_barriers_2021} recognized biological (e.g., age, gender), clinical (e.g., symptom severity), behavioral (e.g., drug/alcohol use), and psychological characteristics (e.g., internal asset). 

However, such models approached mental health help-seeking at the population level, describing the factors in a static sense in separation from young adults' real-world experiences of help-seeking. Our study conducted in-depth interview and delved into the lived experiences of young adults' mental health help-seeking.

\subsection{Ecosystem view of resource use practices of young adults' help-seeking}
Some scholars used qualitative methods and explored how young adults use mental health technologies in their everyday life \cite{c_feasibility_2022, wong_postsecondary_2021}. However, these studies mostly examined mental health help seekers as “users” of a particular technology (e.g., an application or wearable device). 
%Some recent studies began to investigate the distinct roles of resources, albeit primarily focusing on mental health technologies \cite{ralston_fulfilling_2019, stawarz_use_2019}. 
Only a few studies viewed people facing mental health challenges as “persons” surrounded by the dynamic social, technical, economic, and cultural environments and explored their subjective experiences in their everyday lives. For example, \textcite{feuston_everyday_2019} analyzed the narratives about living with mental illness on social media platforms and uncovered how social and technical aspects can marginalize such experiences. 

Some researchers have taken steps to integrate mental health technologies within young adults' social ecosystems and support networks, co-designing technologies that cater to their daily routines \cite{lattie_designing_2020, le_exploring_2021, stefanidi_children_2023}. However,
based on a critical reflection on the previous research, \textcite{burgess_technology_2021} pointed out that the prior studies predominantly focused on one specific platform and argued that future research should take the "\textbf{technology ecosystem}" view to research the experiences of individuals living with mental illness, understanding the entirety and intertwined relationships among different technologies. 

The concept of the "technology ecosystem" illuminated the intricate landscape of digital technology but overlooked resources beyond digital platforms, such as friends and professionals, along with social and cultural factors \cite{ongwere_challenges_2022}.
%An in-depth comprehension of the \textbf{socio-technical ecosystem} and interactions among its components is crucial to identifying opportunities for developing user-friendly technologies. This understanding not only ensures technology adoption but also maximizes the financial and social benefits of investing in these technologies.
To comprehensively grasp young adults' resource utilization practices in mental health help-seeking, we employed the \textbf{Social Ecological Theory} \cite{stokols_translating_1996} to capture behaviors interacting with multilevel resources.
This is a guiding framework prevalent in fields like community health promotion, contends that a holistic understanding necessitates examining multiple levels of analysis \cite{mccloskey_principles_2011}. 
\textcite{mcleroy_ecological_1988} identified five levels influencing human health behavior: individual characteristics, social relationships (e.g., family and friendships), organizational factors, and community characteristics. The societal level was added to the framework to encompass physical, social, and political environments \cite{mccloskey_principles_2011}. These five levels exert a profound influence on individuals' health and well-being \cite{sallis_ecological_2008,trace_information_2023}.

Thus, this study aimed to extend the "technology ecosystem" view \cite{burgess_technology_2021} to a more holistic sociotechnical perspective. We aspired to obtain a comprehensive understanding of what resources they use in mental health help-seeking and how they navigate the ecosystem with diverse resources. 
%incorporate and categorize 

%First contact with health care has significant influences on the following help-seeking intentions. Perceived helpfulness of websites for mental health information \cite{oh_perceived_2009}\cite{gonsalves_design_2019}

%Further, while they offered a structural perspective to examine the streams of retainment, loss, and transformation of potential patients, they failed to uncover the interactions and experiences of patients in contact with each access point, which could shed light on the barriers and facilitators at each center. 

%\textbf{Aims: There are psychiatric models of mental health help-seeking and studies testing the efficacy of single and standalone technologies, Our study aimed to merge the gap between these studies and integrate the role of technology into the behavioral models of help-seeking. }

\section{Methods}
We adopted in-depth interviews to understand the lived experiences of young adults’ mental health help-seeking practices.
\subsection{Recruitment and participants}
We disseminated recruitment messages on the listserv of a southern university in America as previous studies suggested that university students face high and multifaceted pressure from studying and job-seeking \cite{hunt_mental_2010}. We also posted recruitment messages on the social media platform Reddit, as it reaches out to a broader young adult population with more varying backgrounds. Interested participants ages 18-25 were invited to finish a screening questionnaire where we asked for their demographic information, such as age, gender, education, marginalized identities (e.g., first-generation college students, low-income, LGBTQ+, race minority), mental illness diagnosis, help-seeking practices (e.g., previously used resources and technologies such as family, friends, social media, and apps), and an open-ended question about the most recent/impressive mental health help-seeking experience. We also include the PHQ-4 \cite{kroenke_ultra-brief_2009}, a validated screening tool for depression and anxiety to get a rough estimate of their current mental health status. We selected participants based on their responses to the screening questions to maximize the sample diversity.
%It consists of four questions asking about the frequency of feelings of "nervous, anxious or on edge", "not being able to stop or control worrying," "little interest or pleasure in doing things," and "feeling down, depressed, or hopeless" over the last two weeks. 

As shown in Table \ref{tab:participants}, 18 participants were interviewed (7 females, 9 males, and 2 trans/non-binary; 6 White, 5 Black, 4 Asian, and 3 Hispanic). Half of the participants were still in university at the time of the interviews, 6 had graduated from colleges, and 3 had no college education. Thirteen participants had formal diagnoses of mental illness, mainly depression and anxiety. Their anxiety and depression status indicated by PHQ-4 ranged from mild (3-5), moderate (6-8) to severe (9-12). 

\begin{table}[h]
\caption{Participants Information}
\label{tab:participants}
\resizebox{\columnwidth}{!}{%
\begin{tabular}{|l|l|l|l|l|l|l|l|l|}
\hline
\textbf{ID} &
  \textbf{Age} &
  \textbf{Sex} &
  \textbf{Education} &
  \textbf{Race} &
  \textbf{Diagnosis} &
  \textbf{PHQ4} &
  \textbf{Resources Used in Mental Health Help-seeking} \\ \hline
P01 &
  20 &
  F &
  Some college &
  Black &
  depression &
  \cellcolor[HTML]{8AC97D}5 &
  professional help, search engine, friend, family \\ \hline
P02 &
  24 &
  M &
  High school &
  Black &
  depression &
  \cellcolor[HTML]{FFEB84}8 &
  social media, online   communities, friends, professional help \\ \hline
P03 &
  20 &
  F &
  Some college &
  Asian &
  no &
  \cellcolor[HTML]{63BE7B}4 &
  social media, search engine, friends, family, mobile apps \\ \hline
P04 &
  23 &
  M &
  Some college &
  White &
  Anxiety &
  \cellcolor[HTML]{F8696B}12 &
  \begin{tabular}[c]{@{}l@{}}Family, friends, professionals, online communities, services in the local \\ community\end{tabular} \\ \hline
P05 &
  22 &
  M &
  8 through 11 years &
  Black &
  anxiety &
  \cellcolor[HTML]{F8696B}12 &
  family, friend, professional, social media, online communities, telehealth \\ \hline
P06 &
  22 &
  F &
  Some college &
  Asian &
  depression &
  \cellcolor[HTML]{FFEB84}8 &
  telehealth, professional help \\ \hline
P07 &
  21 &
  F &
  Some college &
  Hispanic &
  \begin{tabular}[c]{@{}l@{}}Depression, \\ Anxiety\end{tabular} &
  \cellcolor[HTML]{63BE7B}4 &
  Friends, professors, people with   similar experiences \\ \hline
P08 &
  23 &
  F &
  College graduate &
  Asian &
  Depression &
  \cellcolor[HTML]{D8DF81}7 &
  \begin{tabular}[c]{@{}l@{}}Family, friends, people with similar experiences, Social media, \\ online mental health forums or communities, Search engines, \\ Teletherapy services\end{tabular} \\ \hline
P09 &
  24 &
  M &
  College graduate &
  White &
   &
  \cellcolor[HTML]{8AC97D}5 &
  Family, friends, Reddit \\ \hline
P10 &
  24 &
  M &
  College graduate &
  Asian &
  no &
  \cellcolor[HTML]{63BE7B}4 &
  Hotline, family \\ \hline
P11 &
  22 &
  M &
  Some college &
  White &
  Depression &
  \cellcolor[HTML]{FCAA78}10 &
  Family, professionals, social media, search engine, teletherapy \\ \hline
P12 &
  20 &
  F &
  Some college &
  White &
  Depression &
  \cellcolor[HTML]{F8696B}12 &
  \begin{tabular}[c]{@{}l@{}}Family, professionals, people with similar experiences, social media, \\ online communities, teletherapy, hotlines, search engines\end{tabular} \\ \hline
P13 &
  22 &
  M &
  Some college &
  White &
  Depression &
  \cellcolor[HTML]{FECB7E}9 &
  Family, friends, search engines, mobile applications \\ \hline
P14 &
  21 &
  F &
  College graduate &
  Hispanic &
  Depression &
  \cellcolor[HTML]{8AC97D}5 &
  Professionals, online communities \\ \hline
P15 &
  25 &
  \begin{tabular}[c]{@{}l@{}}Trans\end{tabular} &
  Some college &
  Hispanic &
  \begin{tabular}[c]{@{}l@{}}Depression, PTSD,\\anxiety, ADHD\end{tabular} &
  \cellcolor[HTML]{FA8A72}11 &
  Family, professionals, social media, search engine \\ \hline
P16 &
  24 &
  Trans &
  College graduate &
  Black &
  Depression &
  \cellcolor[HTML]{FFEB84}8 &
  friends, professional, stranger \\ \hline
P17 &
  21 &
  M &
  High school &
  Black & no
   & 
  \cellcolor[HTML]{B1D47F}6 &
  \begin{tabular}[c]{@{}l@{}}Family, friends, professionals, social media, search engine, \\ teletherapy, services in the local community\end{tabular} \\ \hline
P18 &
  22 &
  M &
  College graduate &
  White &
  no &
  \cellcolor[HTML]{FCAA78}10 &
  Family, friends, social media, mobile apps, services in local communities \\ \hline

\end{tabular}%
}
\footnotesize Notes: PTSD: Post-traumatic stress disorder; ADHD: Attention-deficit/hyperactivity disorder
\end{table}

\subsection{Interview Procedure}
The interviews averaged a duration of 1 hour and 20 minutes and each interviewee was compensated with a \$30 Amazon gift card. The interviews took place virtually via Zoom from January to September 2023. Before each interview, we introduced the scope of this study and answered any questions concerning the informed consent form. We also assured the participants that the interviews would be anonymous, confidential, and non-judgmental, encouraging them to be as open as they wanted.

We began the interviews by asking participants' current jobs, residential areas, and other background information. Then, the interviewer asked participants to describe their current mental health status. Example questions included, “Could you tell me about your mental health concerns?” “When did you first notice that?” and “How did it evolve over time?” 

\begin{figure}[h]

% \begin{subfigure}{0.3\textwidth}
% \centering
% \includegraphics[width=\linewidth]{images/Resources.png}
% \centering
% \caption{Visual prompts during interviews}
% \label{fig:visualprompts}
% \end{subfigure}
% \begin{subfigure}{0.6\textwidth}
\includegraphics[width=.8\linewidth]{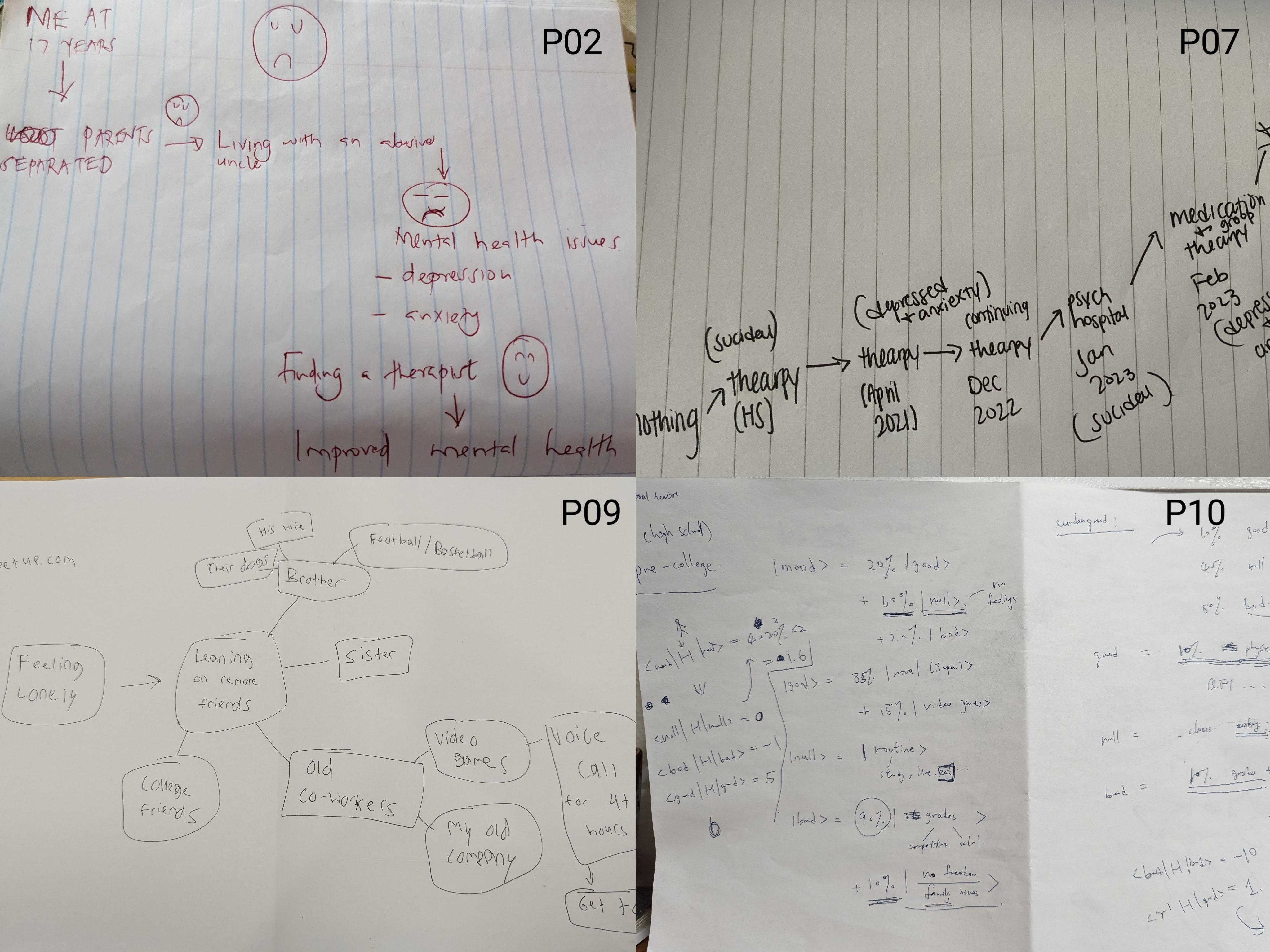}
\caption{Examples of Journey Maps Drawn by Participants. P02 and P07 presented their help-seeking timeline, P09 demonstrated his resources network, and P10 quantified his satisfaction with each resource using self-defined mathematic formulas.}
\footnotesize
\Description{
 Examples of Journey Maps Drawn by Participants. P02 and P07 presented their help-seeking timeline, P09 demonstrated his resources network, and P10 quantified his satisfaction with each resource using self-defined mathematic formulas.}
\label{fig:examplemaps}
% \end{subfigure}

% \caption{Interview materials}
% \label{fig:interviewmaterials}
\end{figure}

We utilized \textbf{visual elicitation techniques} for a richer exploration of their (non-)help-seeking journeys \cite{chen_timeline_2018}. When participants started to dive into specific story-telling, we asked them to draw their experiences of when, where, and how they sought help. Participants were encouraged to take some time to recall their experiences and present their stories in any visualization format. Our participants drew their help-seeking experiences from different perspectives, shown in Fig \ref{fig:examplemaps}. The interviewer then asked follow-up questions based on the drawings to elicit details, for example, “What did you do to cope with [a specific event] in the drawing?” “Did you talk with any people (e.g., family, friends) or use any technology (e.g., social media groups, mobile apps, hotlines)?” and "Why did you decide to do that?" In this process, the interviewer paid special attention to how they perceived and used different resources to seek help. The interviewer also encouraged participants to continually add contexts to the drawings as they narrated. Participants were instructed to focus the camera on the drawing to share with the interviewer during this process. 

After the interview, we asked participants to reflect on their experiences and evaluate their satisfaction with each mentioned resource using a five-point scale (1- very unsatisfied; 5- very satisfied). We then invited them to share the most satisfying and unsatisfying experiences, the challenges they encountered during help-seeking, and the ideal help they wished to get.
% reminded them to keep developing the drawing prompts like "they are great visuals to understand your story. Can you keep drawing as you tell me?" 
%After going through the help-seeking experiences on the journey map, we screen shared a checklist that aggregated resources listed in the U.S. Surgeon General’s Advisory \cite{office_of_the_surgeon_general_osg_protecting_2021} to remind participants of their interactions with other resources. 

\subsection{Ethical concerns}
This study was approved by the University Institutional Review Board (IRB). To protect the safety of our participants, in the informed consent form and at the beginning of interviews, we reminded the participants that they should feel free to take breaks during the interview and that they could exit the interview at any time. During the interview process, we kept sensitive to participants’ emotional changes when difficult experiences were disclosed and checked whether they would like to continue when negative emotions were observed \cite{draucker_developing_2009}. After the interview, we provided a list of mental health resources to the participants for future use, which was considered helpful and appreciated by many participants. 

\subsection{Data Analysis}
All the interviews were conducted by the first author and were audio recorded and transcribed. The interviewer wrote debriefs immediately after each interview. We used iterative open coding and axial coding methods \cite{corbin_basics_2014} to systematically analyze the transcripts. The analysis was assisted by using NVivo, a qualitative content analysis software. This iterative process continued until the twelfth interview, at which point the codes reached a stable state, forming the foundation of our initial open coding schema. The open coding schema includes codes about resource type (e.g., technologies, family, professionals), support type (e.g., informational support, distraction), and challenges (e.g., high-cost, hard to talk). 

\begin{figure}[h]
    \centering
    \includegraphics[width=1\linewidth]{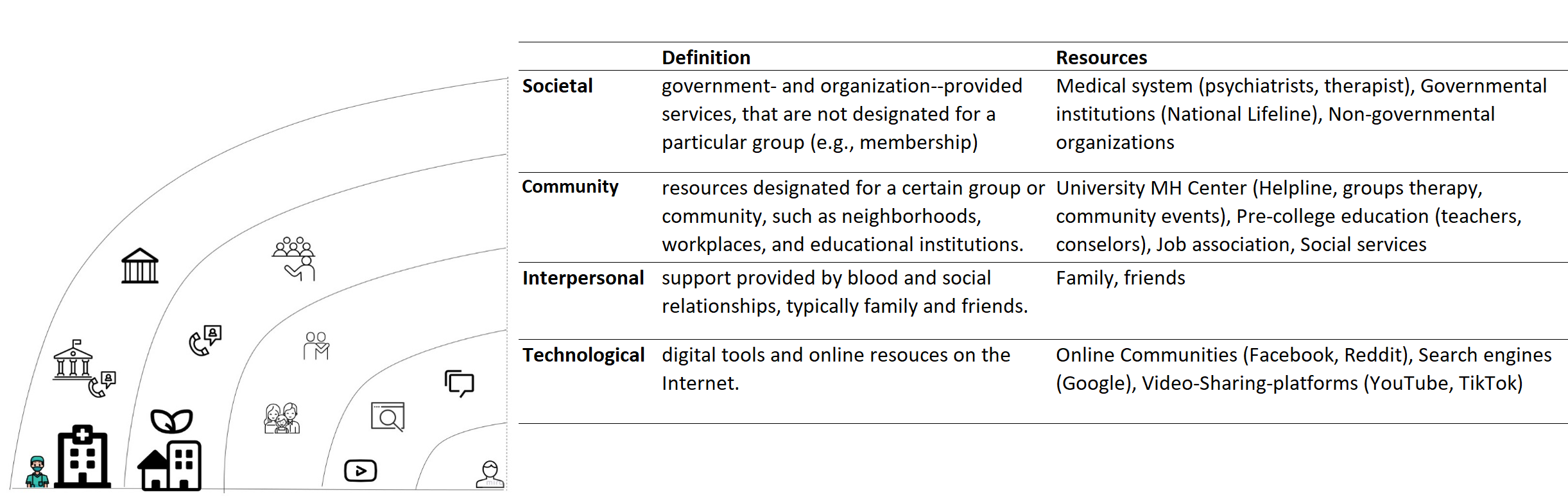}
    \caption{A Four-level Socio-technical Ecosystem Framework of Mental Health Help-seeking Resources}
    \label{fig:ecosystem}
\end{figure}

The participants reported broad resources from which they sought help. We followed the Social Ecological Theory \cite{stokols_translating_1996} 
%to conduct \enquote{ecological analyses characterize environmental settings as having multiple physical, social, and cultural dimensions that can influence a variety of health outcomes.} 
and categorized these resources into four levels: technological, interpersonal, community, and societal levels, forming a \textbf{Socio-technical Ecosystem Framework} of resources as shown in Fig. \ref{fig:ecosystem}. The technological level emerged as a new category through our data analysis, while the other three levels were previously discussed in existing literature \cite{stokols_translating_1996,thompson_social_1990,seligman_depression_1975}.

%We revisited the interviews and mapped the five categories according to the chronological order of participants' narratives. We revised the names of the categories and built an initial model. This model was further developed in the following interviews and analysis process. For example, the evaluation and use activities were not adequately represented in the previous 12th interviews; therefore, we selectively chose participants with behavioral change goals and focused on eliciting their accounts of evaluating and using videos for the following interviews. We also compared the activities with prior empirical studies and models such as the Use and Gratification Theory \cite{katz1973uses} and the schematic model of information seeking \cite{savolainen2006time}.
%such that emergent themes can be compared with previous literature and further explored in subsequent interviews. The researchers will proceed to axial coding by revisiting the open codes and the corresponding transcripts to refine and conceptualize themes into categories. 

Using this framework, we began axial coding to reorganize codes and generate themes and sub-themes related to participants' practices and challenges in seeking help. The two authors convened weekly meetings to review new interview debriefs, reflect on emerging themes in coding, compare codes with existing literature, and document these reflections in notes and memos. Through discussions, we purposely selected the next interviewees to further test and refine the framework. For instance, since societal resources were less frequently mentioned by earlier participants, we deliberately chose participants who utilized societal resources, focusing on how they defined, accessed, and perceived various societal services in comparison to resources at other levels. The initial open code "support type" evolved into the theme "support mechanisms" with three subthemes in axial coding. In addition, we systematically mapped resources used by participants harnessing the Socio-technical Ecosystem Framework, identifying two types of support system, various pathways, and barriers in help-seeking. 

\subsection{Limitations}
Our participants were young adults who were willing to talk about their mental health concerns and help-seeking processes. They might also have less self-stigma. Thus, we acknowledge that their help-seeking behaviors could not represent all young adults, particularly those who did not seek help or were unwilling to talk about their experiences. Future research can examine the non-help-seeking behaviors of young adults.

\section{RQ1: Resource use practices of young adults' mental health help-seeking}
We first described the Socio-technical Ecosystem Framework of resources for young adults' mental health help-seeking consisting of four levels of resources and two types of support systems. Then, we illustrated the mechanisms of how the ecosystem of resources supports young adults and young adults' pathways to care in the ecosystem.

\subsection{A Socio-technical Ecosystem of Resources for Mental Health Help-seeking}
\begin{figure}[h]
    \centering
    \includegraphics[width=\linewidth]{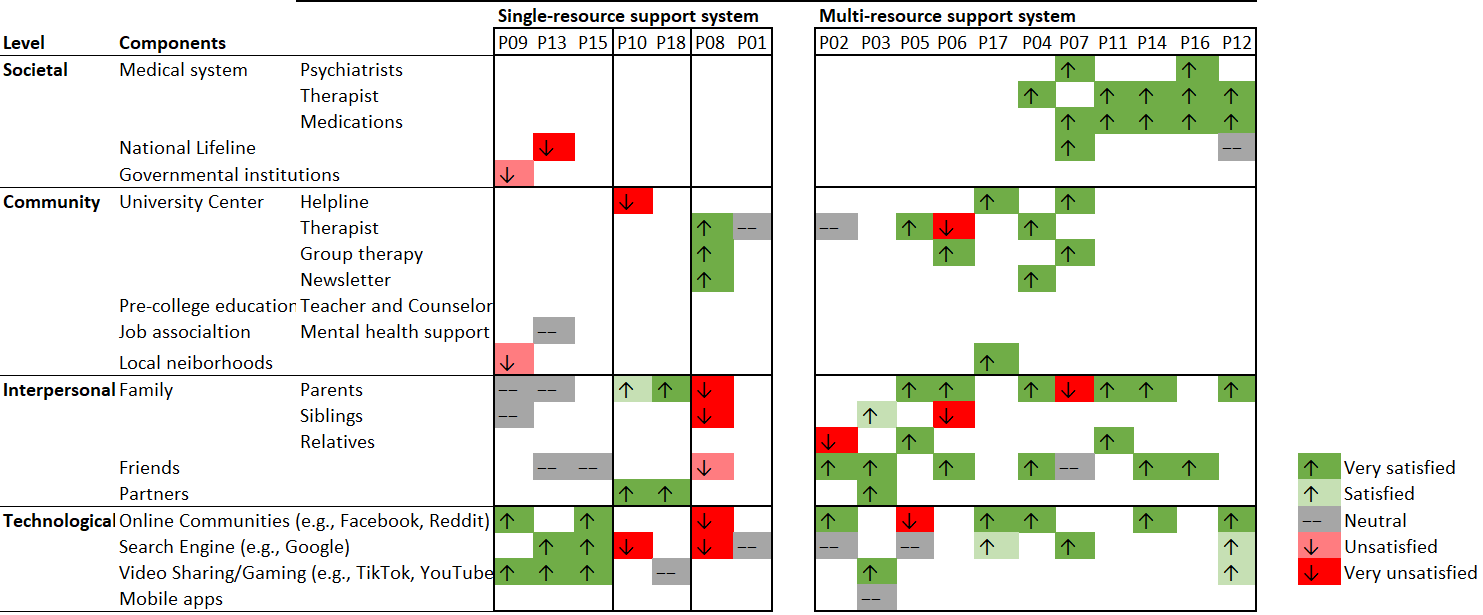}
    \caption{Participants' Support Systems during Mental Health Help-seeking. We incorporated a resource in the participants' support system only if they rated it as "satisfied" or "very satisfied".}
    \label{fig:supportsystem}
\end{figure}

We mapped the resources participants used in mental health help-seeking. We labeled their satisfaction with each resource according to their answers to one interview question: \enquote{To what extent are you satisfied with [a certain resource]?} We included a resource in the participants' support system only if they rated it as \enquote{satisfied} or \enquote{very satisfied}. 
Further, we clustered participants' resource use patterns according to their support system structure as shown in Fig \ref{fig:supportsystem}. Seven participants had a single-level support system where they solely relied on a certain level of resources (e.g., technological level) and eleven participants had a multiple-level support system. 

The \textbf{technological level} includes search engines, online forums, video-sharing platforms, and mobile apps. These resources were frequently mentioned by participants and most of the use experiences were satisfying, but it should be noted that these resources had relatively small effects on participants. 
The \textbf{interpersonal level} consists of participants' interpersonal ties. Except for P01, all other participants at least tried to seek help from family, friends, relatives, or partners, although six of them got neutral or unsatisfying feedback. 
The \textbf{community level} contains all resources provided by entities that participants are members of, such as pre-college education, universities, companies, and local neighborhoods. 11/18 participants tried community-level resources and 6 participants got various extent satisfying experiences. 
The \textbf{societal level} contains non-profit institutions and professional care including professional help from the medical system, the National Lifeline, and governmental institutions. Only 8 participants utilized these facilitates. 

% We did not classify mental health services at the community level as professional, because the professionality of these services varies drastically. 

\subsubsection{Single-resource support system} 
Seven participants relied on one level of resources because of previous failed attempts and low economic status. No participants solely relied on the societal level help. 

As shown in Fig \ref{fig:supportsystem}, they more or less tried other resources but had unsatisfied or natural experiences. For example, P09 contacted the government institutions for unemployment subsidies and tried to find community activities in the new neighborhood, ending up with disappointment. His conversations with family were not very helpful and he ended up relying on \textit{technological} resources such as browsing similar experiences on forums and playing video games. 
The other six participants attempted to expand their resources as well but failed to get the expected outcomes, resulting in the single-resource support system.
Low economic status may also contribute to the single-resource use pattern. Five out of the seven participants came from low-income families so they could not afford professional help at the \textit{societal} level. Some participants (e.g., P15 and P18) did not belong to a community (e.g., university, company) and thus had no access to any community resources.

%Two participants (P01, P08) relied on the university mental health center, where they got the diagnosis, free therapy sessions, and other support. They did not have friends and were unwilling to seek help from technology and family. Almost all participants who were at college/university utilized college/university-provided resources. In P05's case, the gang was the dominant community power manipulating resources, rather than the college. If they did not belong to any community that provided mental health support and had no access to community resources, participants were likely to be blocked by professional help. 

%Interpersonal-supported system could mitigate the need to seek professional help. Among the five participants without an official diagnosis, four of them relied on interpersonal help completely or partially. Four participants (P02, P05, P16) mainly relied on family or friends to deal with mental health concerns. As P02 said, \pquote{I'm open to trying different sources, but at the moment, I can't say specific sources because, at this point, I'm really happy with what I have, because day by day I'm becoming better and better} and the \pquote{hope} given by his friends was all he needed. P16 referred to her friends as her \pquote{support system}.

\subsubsection{Multi-resource support system}
Eleven participants had more than one level of resources to seek help from, which is more robust and stable.

Our participants exhibited various patterns in resource combinations such as technological + interpersonal, interpersonal + community, and technological + interpersonal + societal. Only one participant (P04) utilized all four levels of resources, and three participants (P07, P14, P12) utilized three levels of resources in their help-seeking. The other seven participants mainly relied on two levels of resources. 

Having multiple resources had advantages such as offering backups when needed, getting different support from different sources, and resisting external changes. For example, the strong interpersonal support encouraged P11 and P16 to seek professional help. P04 and P12 strategically sought emotional support from online communities when their therapists and friends were unavailable.
%An ideal support system consists of clinical support and emotional support. The most important thing for them is a reason or a strong motivation to keep up. The motivation to keep seeking help comes from other types of day-to-day interactions that provide social support and other types of help (housing, accommodation, and financing)
%Seeking formal help requires strong "hope" to control the situation and get better, which many times comes from informal social support. If the efficacy of the formal treatment does not go well. In the involvement of community-level help, teachers, and universities should be more sensitive to why young adults prefer to skip this level and perform flexible procedures case by case.

\subsection{Support Mechanisms of Socio-technical Ecosystem of Resources for Mental Health Help-seeking}
The socio-technical ecosystem supports young adults through three mechanisms, including \textit{care-giving}, \textit{care-mediating}, and \textit{care-outreaching}. Most of the resources can help young adults through multiple support mechanisms. 

\subsubsection{\textit{Care-giving} Mechanism} Resources can directly offer young adults informational and emotional support, distraction, and flexibility of partial disclosure.

\textbf{Informational and emotional support} is the most common and direct support mechanism. Participants directly sought answers related to symptoms, potential causes, and guidance on how to cope with mental distress. For example, the participant's family and friends recommended mental health resources, including therapists and hotlines (P13, P18), covered treatment costs (P02, P04), and provided encouragement for ongoing treatment (P12, P14).

\textbf{Distraction} refers to resources that enable the participants to temporarily get away from everyday distress and mental health concerns. 
Technological resources, such as games and online videos, and interpersonal resources were frequently used by participants for this purpose. For example, P13 commented, \pquote{People, laughing, banter, escapism... I think distraction is actually a very important way, at least to amuse ourselves and just to get rid of the stress from life.}

\textbf{Partial disclosure} is a mechanism that allows participants to freely choose how much to disclose to a certain resource. It permits a flexible way to seek help in their "comfort zone". P14 expressed confidence that \pquote{Obviously they [my friends] are gonna support me,} although \pquote{actually I don't want to tell my mental distress.}

\subsubsection{\textit{Care-mediating} Mechanism} Sometimes, a resource acts as a bridge and leads participants to another helpful resource.

\textbf{Direct mediating} refers to a resource directly leading to another one.  
Technologies such as search engines could generate a direct impact on participants by showing them relevant information. Some participants followed the lead of online information to access professionals. For example, three participants (P01, P08, and P13) mentioned using the Psychological Today website to look for therapists that fit their preferences on factors such as gender, education, location, and cost. 

\textbf{Indirect mediating} suggests that situations when mediating effects take place over time by gradually changing participants' knowledge and understanding of resources. For instance, P07 decided to explore medication as a potential solution after repeatedly encountering recommendations from Google search results.
Some participants developed interpersonal relationships within online groups. For instance, P02 formed connections with two online friends through Facebook groups and frequently shared recent experiences with them.

\subsubsection{\textit{Care-outreaching} Mechanism} \textit{Care-outreaching} means that a resource actively reaches out to potential users to provide help, such as services that directly intervene when certain situations emerge.

\textbf{Awareness raising} typically functions through advertising available resources and encouraging help-seeking. 
For example, P08 was attracted to an event about depression awareness by frequent email reminders from the university. She recalled, \pquote{I think it was like halfway through Covid, I received consecutive emails day by day, saying to focus on your mental health. That was when I felt all the stressors were hitting on me all at once and I wondered why didn't I have a break from school? Let me, just, attend one of these Zoom Meetings, and see what they are about. Let me just check these resources out. It was during that time that I realized that I probably need to get some help because these are the symptoms.}

\textbf{Active intervention} refers to resources reach out to potential help-seekers and initiate help-seeking. 
For example, P16, who locked herself at home for two months following her mother's passing without realizing she might have depression, had friends who constantly checked on her and eventually took her to the hospital. She expressed gratitude for their intervention, stating, \pquote{This was too much for me. They came in when I lost it during that time. They took me to the hospital, checked with the hospital, and showed that I really needed this [professional help].}

%To cope with depression and anxiety, P06 also tried to talk with family and friends, seek financial support from her university emergency center, utilize free therapy sessions provided by the university mental health center, and participate in group therapy in research teams. Some of the attempts failed and some of them succeeded. 

\subsection{Pathways to Care in the Socio-technical Ecosystem}
To understand how participants interact with resources in the socio-technical ecosystem for mental health help-seeking, we depicted various pathways to care via the four levels of resources.

\subsubsection{Technological level pathways} 
\begin{figure}[h]
    \centering
    \includegraphics[width=\linewidth]{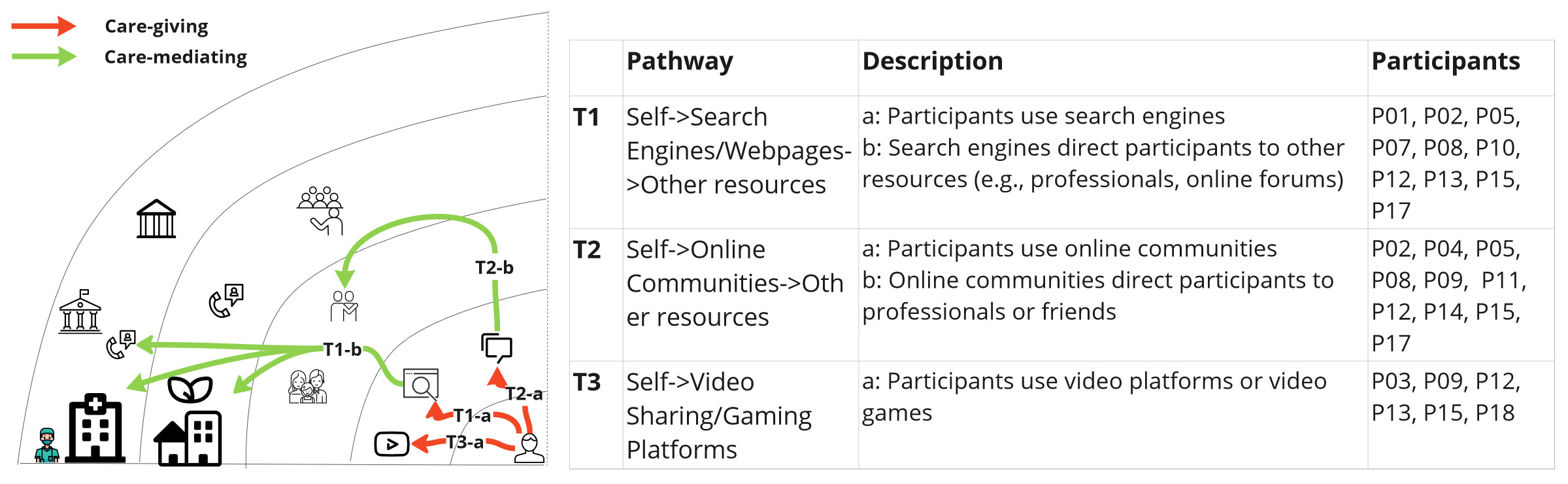}
    \caption{Technology level pathways}
    \label{fig:technology}
\end{figure}

Fig \ref{fig:technology} shows the pathways to care through technologies.

%Self->Search Engines/Webpages->Professionals/online forums
%(T1) demonstrates how participants mostly got informational support from online information from search engines, online forums, and video-sharing platforms. 
Participants used search engines and websites (T1-a, Fig. \ref{fig:technology}) to search for information about symptoms and treatment. 
On online communities (T2, Fig. \ref{fig:technology}), participants read other people's stories, commented on posts, and shared their own experiences. P05 and P18 considered "online friends" even better than their "real friends." P12 explained, \pquote{I will say this online group is safer because the online friends have seen their issues but, the majority of my physical friends don't experience what I'm experiencing.}

%Self->Video Sharing/Gaming Platforms
Turning to online video platforms (T3) was another frequent pathway and acts as a \textit{care-giving} resource. Some participants watched videos for its distraction mechanism. For example, P03 intentionally immersed herself in TikTok videos to relax and P13 watched YouTube videos to escape from mental health worries. Other participants also used video platforms similar to search engines for informational support.

\subsubsection{Interpersonal level pathway} 
\begin{figure}[h]
    \centering
    \includegraphics[width=\linewidth]{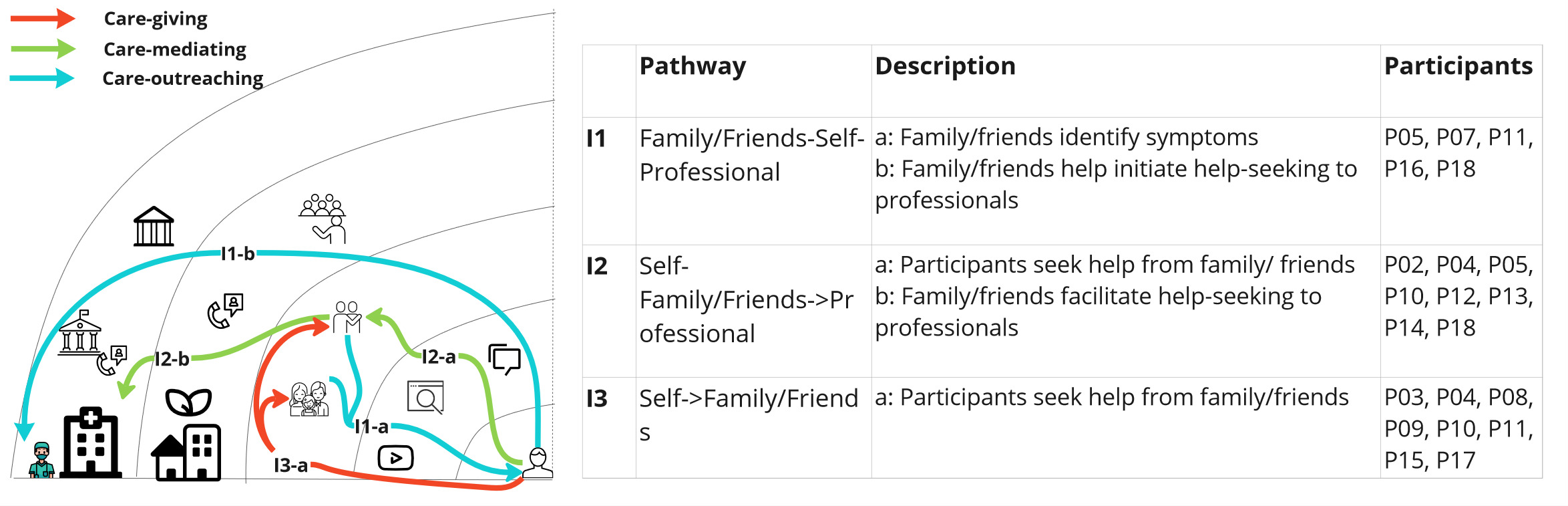}
    \caption{Interpersonal level pathways}
    \label{fig:Interpersonal}
\end{figure}

Interpersonal relationships contain the most natural sources of help- blood and social relationships, typically family and friends. 
Figure \ref{fig:Interpersonal} shows the pathways at this level.

Family/Friends-Self-Professional (I1) is a reflection of \textit{care-outreaching} mechanism at the interpersonal level. This pathway was mentioned by five participants. Family and friends play a crucial role in identifying early-stage symptoms due to their daily interactions with the individuals, especially when they were during their childhood. P07 and P16 were taken to psychiatrists by parents at elementary schools.

Six participants openly shared their mental distress with interpersonal relationships (I2-a). I2 represented how family and friends directly mediated their help-seeking process (I2-b). P02's family and P05's friends provided financial support to cover their therapy expenses; P10 and P13's family accompanied them to contact professional help. 

As illustrated by Self->Family/Friends (I3), interpersonal relationships could be a source of \textit{care-giving}. Some participants found solace in the companionship of family and friends and used it as a substitute for professional help. For instance, P18 mentioned that \pquote{she came with immediate effects. She encouraged me, talked to me, and we played games.}

\subsubsection{Community level pathway} 
\begin{figure} [h]
    \centering
    \includegraphics[width=\linewidth]{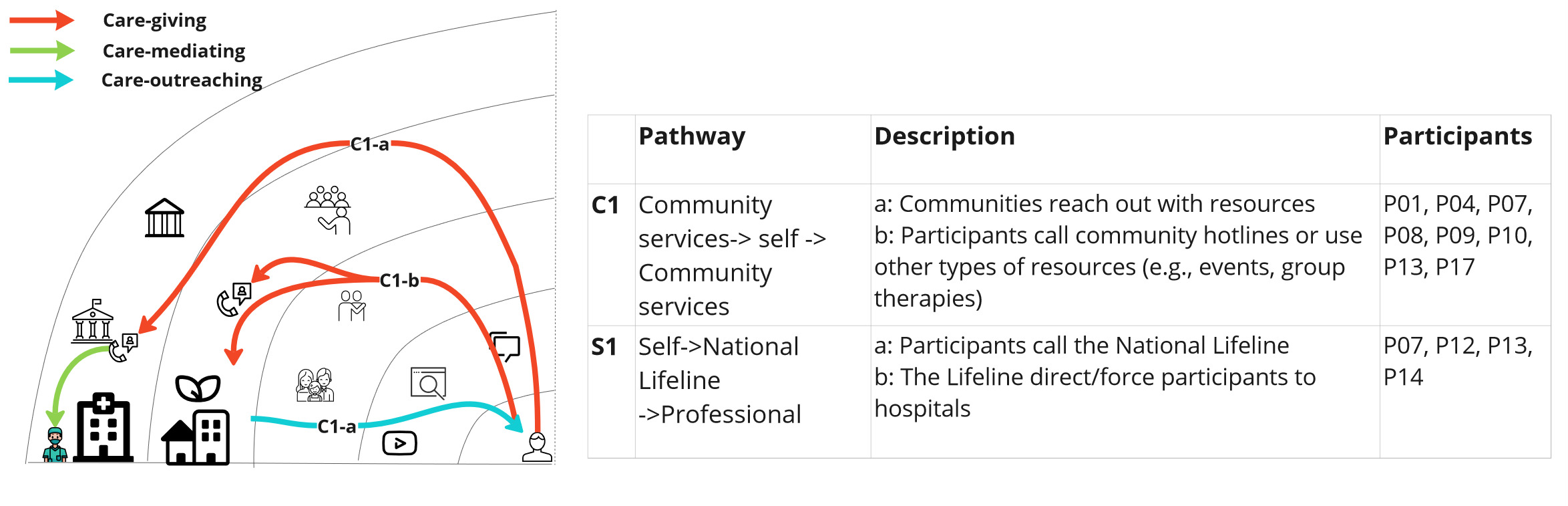}
    \caption{Community level pathways and Societal level pathways}
    \label{fig:Community}
\end{figure}

Community mental health resources were primarily accessible within educational institutions, including schools and universities. In contrast, entities like residential areas and commercial companies may lack investment in mental health services. For instance, P13, a sports official, highlighted limited workplace-based mental health assistance, indicating disparities in stress management training.
Fig \ref{fig:Community} shows the pathways at this level. 

%preliminary, middle, and high schools. Before university, many schools do not set up an independent department with professionals in mental health, thus teacher is an available source perceived by adolescents. 
%Community services->self -> Community. 
Some universities played essential roles in proactive \textit{care-outreaching} to young adults (C1-a). They function by the awareness-raising mechanism, specifically, disseminating information and resources through diverse channels, including emails, social media platforms, and awareness events. P10 and P13 noted the pervasive presence of university hotline signs in hallways, offices, and dining halls. 

In C2-b, community services had advantages in fulfilling the \textit{care-giving} mechanism. Communities aggregate multifaceted resources and a set of comprehensive policies around mental health accommodations, which are important incentives for participants to seek help from community services. 
Within university settings, community mental health support was effectively integrated with other accommodations (C1-b). P06 requested a reduced class load, P07 received extended tuition payment options, and P08 benefited from continued free therapy sessions post-graduation. They heavily relied on universities' free resources (e.g., psychiatry visits and free therapy sessions).

%P07 shared her experience seeking help from one of her teachers, \pquote{I remember I cried to her about how I was feeling, and it was just nice to let someone at school know how it's feeling. I felt like I could breathe better with her knowing.} 

\subsubsection{Societal level pathway}
%includes access to professional help through national lifeline, governmental institutions, and non-governmental organizations.
%Direct pathway to care illustrated that participants got in touch with professional mental health services directly, including hierarchical institutions and health workers such as hospitals, clinics, psychiatrists, therapists, nurses, etc. 
No participants directly accessed professional help, or successfully used governmental institutions or non-governmental organizations to access professional help. The barriers in such pathways are discussed in Section 5.
% 1. self->national lifeline (as a caregiving source)
% 2. Self->national lifeline -> professionals (as a mediating source)

Self-National Lifeline-Professionals (S1) was mentioned by several participants as a channel to turn to in an emergency (S1-a, Fig \ref{fig:Community}). P07 called this lifeline after she cut herself but her therapist was unavailable. She commented, \pquote{I just wanted to talk to someone that would listen to me... I think I got what I needed out of it, like just talking to someone because it was able to calm me down quite a bit.} In some occasions, calling the Lifeline can result in mandatory hospitalization (S1-b). P14 was afraid to use the Lifeline because her friend was placed in a hospital after calling.
%Some participants mentioned their and others' bad experiences with Lifeline because of its limited availability, certain mandatory policies, and disappointing responses. 
%P14 called Lifeline in 2018 which put her on hold for two hours and did not go through. P13 was discouraged by the staff who answered his call because \pquote{I wasn't really actively suicidal right then, but I was very much considering it, so I called it. When they realized that I wasn't actively trying to do something, they kind of seem to more blow me off, as if oh everyone has bad days, yeah, you're depressed, but you know, other people have it worse. And I was like, oh, okay. Thank you.} 

\subsubsection{Cross-level pathway} \label{para: multi} We identified two pathways that involve close collaborations across multiple levels of resources.
\begin{figure}[h]
    \centering
    \includegraphics[width=1\linewidth]{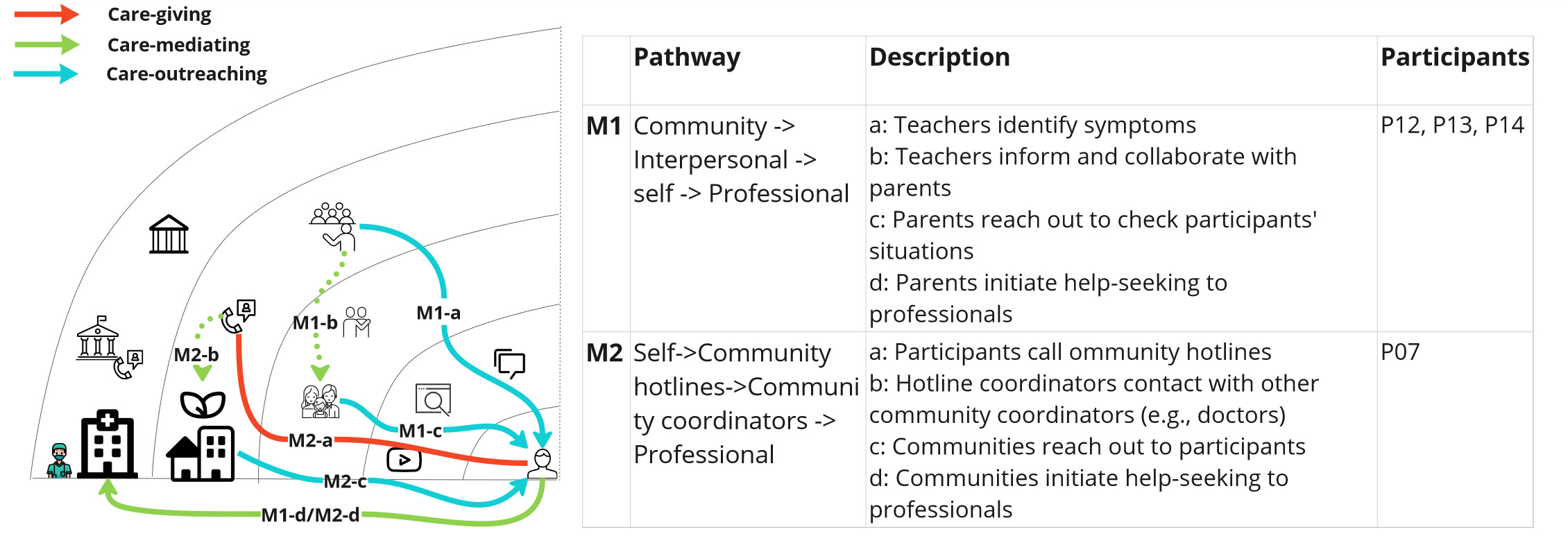}
    \caption{Cross-level pathways}
    \label{fig:multiple}
\end{figure}
%How to further abstract the pathways below? they are very empirical. 
% community -> interpersonal -> self -> professional 
% self->community hotlines->community coordinators -> professional <should this belong to self->community sources->professionals, given both hotlines and coordinators are community sources>

Community-Parents-Self-Professionals (M1) demonstrated teachers pivotal role in recognizing and engaging parents in addressing mental health symptoms during participants' youth. For three participants, teachers identified signs of mental health issues and informed their parents, leading to professional intervention. In addition to notifying parents, teachers monitored participants' recovery and facilitated timely information exchange with parents.
%\pquote{In senior year of high school, they [suicidal thoughts] got really bad and I went to see one of the counselors.} 

Self-Community Hotlines-Community Coordinators-Professionals (M2) illustrated the effectiveness of the community in responding to emergencies. P07 called the university crisis line to request accommodations. Upon learning about her self-harm tendencies, the staff dispatched a police officer and a doctor to her residence. They escorted her to a local hospital, where she was diagnosed with major depressive disorder and generalized anxiety disorder and got medications, as she recounted,\pquote{I stayed in the hospital for two days..they diagnosed me with a major depressive disorder and generalized anxiety disorder, and then that's when they gave me a medication to start to treat both of those.}

\section{RQ2: Challenges of young adults' mental health help-seeking}
\label{sec: challenges}
The Socio-technical Ecosystem Framework enabled a thorough analysis of pathways, facilitating the identification of specific barriers encountered by our participants at each resource level, and distinctive challenges young adults confront in establishing a robust support system.

\subsection{Barriers to mental health help-seeking at each level}
We first summarized the barriers to seeking mental health help from resources at each level.
%in Fig \ref{fig:challenges}.
 % \begin{figure}[h]
 %     \centering
 %     \includegraphics[width=.75\linewidth]{images/Challenges.png}
 %     \caption{A summary of barriers and facilitators at each level}
 %     \label{fig:challenges}
 % \end{figure}
%Among the ten participants with available families or friends, five (P02, P03, P05, P06, P10) were willing to disclose their mental health concerns to some extent and five were not (P01, P09, P13, P14, P16). 

\subsubsection{Technological level: Limited \textit{care-giving} and \textit{care-mediating} efficacy} 
Although widely used by our participants, technological-level resources seemed only to play an auxiliary role.

\textbf{Auxiliary role in \textit{care-giving}.} 
Participants commonly turned to technology as their initial source for seeking help, utilizing it as an extension of self-coping through activities like accessing online information and finding relatable experiences. Despite offering a sense of anonymity (e.g., learning knowledge without telling others), technology's role remained primarily supplementary for most participants. Notably, four individuals who attempted to rely solely on technology for support ultimately shifted to other help-seeking levels. Additionally, some participants, such as P06 and P10, explicitly expressed aversion to using technology for mental health challenges, because \pquote{The search engines, these, online things are too general. They are helpful when you search for a road trip, but it's not the mental health thing} (P10). Three other participants rarely engaged with platforms like social media or online forums in general.

\textbf{Inadequate mediating effect to societal and community resources.} 
Many participants tried to use technology as a mediator to find other levels of resources for help. However, many resources web pages provided proved to be inaccessible. For instance, P15, unable to afford professional help, attempted to look for free therapies from the government and communities, only to find the process frustrating and challenging. She expressed, \pquote{every time I've looked it up\elide{. Maybe I was just not in a good state of mind, but} it seemed inaccessible. Not every website is very user-friendly. You have to go digging and digging to find what you need to find. And when I'm frustrated and feeling like I need help, that's the last thing I want to do.} She speculated that these web pages might intentionally be designed to be inaccessible, potentially driven by profit motives, stating, \pquote{Honestly, I've studied how Google search comes up. Results from Google are not mainly from your local clinics. It's people that pay more money for their website to show up first...But they just look like businesses.} This inaccessibility, coupled with underlying socioeconomic barriers, extinguished her hope of finding community- and societal-level resources.

%Technologies bring online information about the available resources at societal and community levels, but most participants would not seek further help.

\textbf{Emotional frustration in solitary technology use.}
While technology promises convenience and privacy, the experience of seeking support becomes a solitary journey, often leaving emerging problems unaddressed. A common issue is the overwhelming feeling of sifting through a vast amount of information. For instance, P15, anticipating emotional support from Facebook Groups, expressed frustration: \pquote{having to dig for it can be a little bit demoralizing. I just want instant gratification, but when your brain's all cluttered, the last thing you really want is to have to fight for help.} The struggle to find information not only adds emotional burden but also, as described by P08, intensifies anxiety: \pquote{I get more worried when I look for it because there's just so much information out there... this information on Google might just be like for everybody. I get more anxious because I can't find information that is tailored to my needs.} The lack of personalization can be particularly detrimental to marginalized individuals, making it harder for them to find relatable experiences. For instance, P05, encountering predominantly unrelated experiences about anxiety, felt marginalized and further neglected: \pquote{I'm sorry, mine [situation] was going up for two or three weeks. Most of them are like coincidences. So I feel like I was tragic and traumatizing because I was being attacked and my friend was beaten to death.}

%\pquote{I just kinda, like, wanna keep it private. But at the same time, I know that people use it as a safe forum to discuss their issues and get comments and feedback. But I don't think it occurred to me at all to post my question. I don't feel comfortable releasing it to the public} (P08)

\subsubsection{Interpersonal level: Difficulties in self-disclosure} Navigating interpersonal communications in help-seeking poses a complex challenge for many participants.

\textbf{Lack of channels and skills of self-disclosure.} Six participants expressed a desire to seek help from family and friends but faced challenges in effectively communicating their vulnerabilities. Initiating discussions about mental health, particularly in close relationships, proved to be a common obstacle. P13 commented, \pquote{I mean just it's hard to talk about it...I'm not good at talking to people about the hard stuff.}
Four participants perceived that disclosing their situations could burden their loved ones, hindering their willingness to open up. For example, P14 shared, \pquote{I feel like when you start feeling [suicidal], it's scary because I didn't want to tell my parents that I was feeling suicidal. Because once you tell someone that, it's like it becomes really serious. It gets like, they're like what's wrong? Like it's too much.}

\textbf{Lack of a mediator in complicated interpersonal dynamics.} The absence of a mediator in intricate interpersonal dynamics discourages participants from seeking support. Many participants grapple with challenges in initiating open communication, particularly concerning emotional disclosure within intimate relationships, notably with parents. In the context of established family dynamics, P08 shared, \pquote{I think it's perhaps personality, or how I was raised. I'm not big on sharing my vulnerability in front of people. It's like under a fake name.} 
P07, seeking appropriate grieving after her grandfather's death, encountered reluctance or incapacity from her parents to discuss the topic with her. Specifically, P08 hoped she had a mediator, such as a relative familiar with the family dynamics, who could facilitate communication between her and her parents regarding the impact of their conflicts on her. %Many participants identified strained relationships with family or friends as a major stressor in their lives.

\subsubsection{Community level: Risks of unprofessional and temporary help} Community-level resources are not part of the professional medical system and predominantly offer relatively short-time support.

\textbf{Limited professionalism and scope.} 
Community-level mental health resources exhibit considerable variability in the support they offer. Among the resources mentioned by our participants, university services stand out, showcasing heterogeneous levels of professionalism in terms of the range of services, insurance plans, and policies. For example, P15's university provides free and unlimited therapy sessions, while P07's university imposes a three-session limit. Disparities also manifest in outreach efforts, with P05's university demonstrating less proactive engagement compared to the proactive mental health awareness initiatives undertaken by P10's university.

\textbf{Temporary services as transitional hubs.} 
University mental health services often function as transitional hubs between students' prior educational environments and their future professional endeavors. While some universities may have in-house professionals, their collaboration with the broader medical system is typically limited. This limitation can pose challenges to ensuring sustainable care for individuals transitioning from university mental health services to external institutions. P01 highlighted this difficulty, stating, \pquote{It can be hard when you go from university to an external institution because you have to find one that matches your insurance. The first time, I called at least 10 or so people before I got the practitioner. They were like, 'We are not taking patients or we're not taking certain types of patients.' The whole process can be lengthy and challenging}.

\subsubsection{Societal level: Limited accessibility}
%feeling of exclusion, 
Only six participants successfully sought help from societal resources. Participants (e.g., P07, P08, P15) considered that the constrained opening hours of professional help contributed to low accessibility. Further, beyond service hours, seeking societal resources demands extensive preparation, requiring a significant investment of time and effort. Considerations such as in-network insurance, finding suitable time slots, and selecting therapists contribute to the challenges and strengthen the perception of inaccessibility. Participants expressed frustration with the difficulty of finding a compatible therapist. For instance, P01 discontinued therapy due to perceived misalignment, stating, \pquote{She kept talking about sadness and stuff. And I tried to tell her, I'm not sad, I'm tired. I stopped seeing her because I felt like she wasn't really listening to me but just reacting.}
%The "official procedure" involved in seeking professional assistance also contributes to stigmatization, as articulated by P13, who described the difficulties , \pquote{And underneath the bridge over here is where all the mental health and all that building is. They have an office in here. But to get in there, you have to use this entrance, and this is the entrance, and these are the windows of our building.}
%\pquote{the first one I had was just really weird and would like tell me her personal problems.  10:33:41 And I'd be like, oh, okay. And then the second one. Was like, I wouldn't say she was like.  10:33:51 Weird or anything but she wasn't helpful. I'd be like I'm really anxious she like Jocelyn will.  10:33:56 So stop being anxious. You, you just need a stop. Yes. Like I'm not kidding.  10:34:03 She said those exact words. }
% Except for calling the National Lifeline, none of the participants directly sought professional help at the societal level. 
%The roles and structures of resources at the societal level and the community level are similar. They both provide professional help. For participants with the intention to seek professional help, those who were aware of community resources would first go to the community resources. 

\subsection{Challenges in finding a support system}
Beyond the barriers associated with seeking help from resources at a specific level, young adults encounter border challenges that complicate the help-seeking process.

\subsubsection{Socio-economic and cultural backgrounds}
Individuals' reliance on family and friends for support is significantly influenced by cultural background and family dynamics. Notably, participants facing low socio-economic status, such as homelessness or low income (e.g., P05 and P18), lack community-level resources in their socio-technical ecosystems. Additionally, three Hispanic participants (P07, P14, P15) and two Asian participants (P08, P10) intentionally excluded interpersonal-level resources due to cultural stigmas associated with mental health conditions.
%Participants with strained familial relationships often lean towards community-based services, with technology-based services being their last resort. In this context, technology can function as a crucial firewall, directing them to emergency lines when needed.

%\paragraph{Order of resources.} For instance, when diagnosed with depression, P08 initially attempted to seek help from interpersonal and technology-based resources. However, the former was unavailable due to busy schedules, and the latter proved unsatisfactory. She expressed, \pquote{In that situation, because you can't talk with your parents and friends, and talk with anyone, you can only Google this information, but they are not helpful... I really need someone to talk with.} Subsequently, she turned to the mental health center at her university. \pquote{I was aware of the services [in the city]. If the university [resource] was not available, I had to use these resources outside. But I never had to use them, because I can use the ones at the university.}

\subsubsection{Resilience in help-seeking}
In the face of mental health challenges, several participants exhibited remarkable resilience. For example, P03 framed mental health as a personal responsibility for long-term well-being, stating, \pquote[P03]{I don't seek out professionals purely because I know that I'm going to deal with this for the rest of my life. If I'm dependent on a professional to help me, then how am I going to thrive?} P13 also relied on such resilience and developed a routine for managing suicidal thoughts.
%Such resilience will ,  but acknowledged the necessity of professional intervention due to the escalating severity of their mental health challenges.

\subsubsection{Prior negative experiences} Whether or not to seek help for mental health concerns is not a static, simple, and one-time decision, it is dynamic and closely related to the perceived effectiveness of prior help-seeking experiences. For many participants, a lack of confidence in external assistance led them to cope independently or rely on previously satisfactory resources. Notably, five out of seven participants with a single-resource support system recounted unsatisfactory experiences and feelings of "rejection" when attempting to access other levels of resources.
P08 expressed this sentiment, stating, \pquote{I always wish things were different, but I can't change anything. I feel like I was not supported. I was not helped. I was helpless, and no one was there to come for me.} These negative encounters contribute to hesitancy and reluctance to explore alternative avenues for mental health support.

\subsubsection{External changes in young adulthood}
Young adulthood is a period of significant life transitions, such as entering universities, completing education, and relocating for work, introducing external factors that reshape young adults' relationships with various levels of resources in the socio-technical ecosystem. These changes can exacerbate mental health concerns, leading to heightened stress and anxiety. For instance, entering a new environment often means leaving behind existing support networks, as expressed by P17's experience in a different high school, \pquote{It was when the lack of communication came in... the doctor recommended some hotline numbers just in case my therapist was not around. but some people talk rudely they might not give you the maximum attention or treatment.} Four participants, who were previously dependent on university resources, had to seek societal-level outpatient institutions after graduating.

%External factors compass significant and fundamental environmental life transitions, such as entering universities, completing education, and relocating for work. External factors interplay with other challenges by introducing substantial changes to the physical environment, financial situations, and psychological status, reshaping participants' relationships with different levels of resources in the socio-technical ecosystem. External changes can bring in new stressors, create feelings of isolation and loneliness, exacerbate mental health concerns among young adults, and lead to heightened stress and anxiety as one navigates new responsibilities and expectations, further worsening mental well-being. Specifically, entering a new environment involves leaving behind an existing support network. These external changes, while impactful, also introduce new stressors and feelings of isolation among young adults.

%Recognizing the role of external factors in mental health outcomes highlights the importance of addressing these influences to support the well-being of young adults during this critical life stage.

% \begin{figure}[h]
%     \centering
%     \includegraphics[width=.9\linewidth]{images/P14 long-term.jpg}
%     \caption{The changing perceived socio-technical ecosystem of P14}
%     \label{fig:P14long}
% \end{figure}

\section{Discussion}
By investigating the lived experiences of young adults, we proposed a framework informed by the Social Ecological Theory. The socio-technical ecosystem framework of resources for mental health help-seeking consists of four levels of resources, two types of support systems, three mechanisms, and various pathways to care. We further pinpoint the barriers to seeking help at each level and the special challenges faced by young adults. 
We discussed the theoretical contributions and design implications for %the development of effective interventions to raise awareness of treatments and intentions to access mental health services that address the identified barriers and facilitators of young adults’ engagement with mental health information. 
future action research to incorporate technologies into a collective effort to empower vulnerable populations and foster technologies for social good. 

\subsection{The Socio-Technical Ecosystem of Young Adults' Help-Seeking}
Extending \textcite{burgess_technology_2021}'s efforts to investigate the ecosystem of mental health resources, we proposed the socio-technical ecosystem as a framework to systematically investigate young adults' mental health help-seeking practices. 
%Additionally, we align with recent discussions emphasizing the ecological perspective to improve the accessibility of mental health services \cite{ongwere_challenges_2022}.
Under the guidance of Social Ecological Theory \cite{stokols_translating_1996}, we specified the scope of societal-, community-, and interpersonal level resources for mental health help-seeking, and added the technological level to highlight the current practices and importance of digital technologies in mental health help-seeking \cite{johnson_technology-based_2022}. This socio-technical ecosystem framework offers a comprehensive taxonomy to characterize resource use practices and provides future studies with a practical tool to analyze help-seeking experiences of young adults and other populations.

In contrast to earlier models of mental health help-seeking, our four-level framework not only offers a hierarchy of resources but also indicates a preferred order for accessing these resources. While \textcite{rickwood_conceptual_2012} categorized resources based on the formality and professionalism of services, our study reveals that young adults, in their everyday experiences, prioritize the convenience and proximity of resources. In alignment with previous research \cite{lattie_designing_2020,martinez-hernaez_non-professional-help-seeking_2014}, we observed that young adults often turn to non-professional help initially, as these resources are more readily accessible. For instance, a majority of our participants initially sought information through technology, followed by sharing their concerns with family, friends, or university contacts before considering professional help at hospitals. Our framework conceptualizes and visualizes this practice by depicting their socio-technical ecosystem of resources. 

By intersecting four resource levels and three mechanisms, we have outlined potentially essential research topics in mental health help-seeking and discussed the implications of this framework below.
%This framework lays a structured foundation for conceptualizing future empirical studies in the field. 

% \cite{lattie_designing_2020} contextualized the needs of college students in their social ecosystem and social support networks and co-designed technologies that can be fitted into their everyday lives.  \textcite{le_exploring_2021} proposed a multifaceted approach incorporating mobile applications, individual interventions, and naturalistic conversations to mitigate college students' everyday anxiety. 
% Notably, a recent workshop emphasized an ecological perspective to enhance the accessibility of mental health services \cite{ongwere_challenges_2022}.
% <Conceptual framework for personal recovery in mental health: systematic review and narrative synthesis> The roles of pathways and resources \cite{leamy_conceptual_2011}.

%\input{tables/crosstable}

\subsubsection{\textbf{Technological level}}
We added the technological level in the framework which is less represented in Social Ecological Theory \cite{stokols_translating_1996} and previous behavioral models of mental health help-seeking such as \cite{rickwood_conceptual_2012, biddle_explaining_2007}. 

\textbf{Design for technologies' \textit{care-mediating} and \textit{care-outreaching} mechanisms.} Designing self-management tools for mental well-being has been a focal point in prior research \cite{pinto_avatar-based_2013}. However, our participants expressed a desire for technology to connect them to resources at other levels beyond \textit{care-giving}. We identified limitations in current technologies' ability to direct participants to useful resources in Section \ref{sec: challenges}. We advocate for future research to explore the design of technologies that better support \textit{care-mediating} and \textit{care-outreaching} mechanisms, leveraging technology's advantages in availability \cite{pendse_can_2021} and privacy \cite{petelka_being_2020}, and ultimately promoting the willingness to seek professional help. Such practices also align with the increasing attention on the ecological validity of health technologies, emphasizing the need for long-term deployment, tracking, and measures of behavioral changes \cite{sallis_ecological_2008, mcleroy_ecological_1988, fadhil_assistive_2019}. For more specific design implications for mental health technologies, refer to Section \ref{sec: design}.
 
%Our framework categorizes the technology as a distinct level for abstract purposes, but technologies are now pervasive and have been an integral part of social infrastructure. For example, the National Crisis Line and websites of medical systems are integral parts of societal services. Future studies can explore how to maximize the efficacy of technologies in emergencies. 

\textbf{Consider risks of technology as a solution to mental health.} Digital health and mental health technologies have emerged as key research areas in HCI and CSCW, aiming to provide innovative technological solutions to health problems. However, findings from our study and prior studies \cite{gitlow_how_2019} indicate that these tools are less utilized in the lived experiences of young adults. In our study, participants primarily turned to technologies for basic knowledge and emotional preparation before seeking further help from resources at other levels. Notably, three participants solely relied on technologies due to a lack of other available resources (as illustrated in Figure \ref{fig:supportsystem}). The major goal of designing self-management tools has been to encourage self-coping with mental health, but there are risks associated with potentially hindering further help-seeking. Future studies should reflect on the design goals of such apps and consider how to balance self-help and help-seeking dynamics \cite{pendse_treatment_2022}.

%featuring various technology modalities, such as online video \cite{liu_modeling_2023} and virtual reality \cite{xinyue_sally_social_2023}, to directly provide care. 
%However, the credibility evaluation of online health information remains a challenge for participants \cite{liu_consumer_2023}. 

\subsubsection{\textbf{Interpersonal level.}} 
The interpersonal level highlights the significance of peer and family support which aligns with the informal help category in \textcite{rickwood_conceptual_2012}'s model.

\textbf{Facilitate collaborative mental health help-seeking.} Our findings elucidate how interpersonal-level resources provide support through mechanisms including \textit{care-giving}, \textit{care-mediating}, and \textit{care-outreaching}, delineating various pathways that elaborate on the sustained motivation fostered by supportive accountability \cite{mohr_supportive_2011} to seek help and, more importantly, professional treatment. Collaborative mental health help-seeking, as we discovered, necessitates mental health literacy in interpersonal relationships. For example, P14 emphasized the growing capacity of her mother to assist her. To enhance caregivers' mental health literacy \cite{chovil_engaging_2010}, there is an imperative need to develop educational programs focusing on self-disclosure and mental health-related conversations within intimate relationships. 
%Some studies have initiated exploration of interventions such as peer-delivered and technology-supported self-management \cite{fortuna_certified_2018}.

\textbf{Support long-term interpersonal relationship building.} Our findings emphasize that the quality of interpersonal relationships is a crucial prerequisite for fostering self-disclosure \cite{calear_sources_2022}, impacting not only the context of help-seeking but also the overall mental well-being of young adults \cite{meyerhoff_meeting_2022}. Participants noted that family dynamics and cultural norms constrained daily conversations and hindered the communication of negative emotions and open discussions about mental health experiences. Those who benefit from interpersonal help often engage in frequent and open communication, as exemplified by P06 and their families. Challenges in establishing and maintaining friendships and connections \cite{lynch_young_2018} are common for young adults, contributing to a sense of isolation and helplessness when grappling with mental health challenges \cite{li_experiences_2021}. While current research predominantly focuses on help-seeking, future investigations should explore ways to enhance interpersonal relationships in the long term.
%intimacy, connectedness, and trust are crucial prerequisites for fostering self-disclosure \cite{calear_sources_2022}

\subsubsection{\textbf{Community level}} 
Community's role in mental health help-seeking is less investigated by previous work in CSCW. Our framework acknowledges and underscores the pivotal role of community-level resources, particularly within educational environments like colleges and universities.

\textbf{Emphasize University's support for young adults' help-seeking.} 
Universities play a pivotal role in aggregating resources and facilitating young adults' help-seeking. This study reveals that educational institutions, as major settings where young adults spend significant time, serve as crucial sources of support for mental health concerns \cite{laws_students_2013}. A third of the participants in our study predominantly utilized university mental health resources, relying on them for various needs. The university emerges as a key player in raising mental health awareness, disseminating resource information, and offering accessible treatment options. Given that many mental health concerns originate from the university environment, positioning universities as major hubs and providing comprehensive financial and academic accommodations can support young adults more effectively. We advance future research to harness a community-centered approach to explore and implement designs for mental health support within educational and communal settings \cite{liang_embracing_2021}. 
%jointly used website at the community level \cite{williams_impact_2021}

\textbf{Aggregate and mediate resources in the Socio-technical Ecosystem.} 
Previous research recognizes the community's role as a resource hub that aggregates and bridges resources from technological, interpersonal, and societal levels. For example, \cite{cornish_meeting_2017} experimented with a community-level stepped care model to facilitate college students' help-seeking. Gaps were identified in the transition from the community to societal-level resources, with only one participant (P07) effectively introduced to hospital professionals, while others faced obstacles in this transformation. Future researchers are encouraged to explore community-level interventions, including the integration of technologies into care practices \cite{brown_how_2020}. We also discovered the advantages of communities in \textit{care-outreaching}, investigating how to actively reach out to young adults facing difficulties, particularly those lacking interpersonal support. Incorporating lightweight services such as short-message has proven to be a promising direction \cite{rohricht_simple_2021}. 
A significant challenge at the community level is the lack of adequate mental health resources. Strengthening the \textit{care-giving} mechanism necessitates exploration of the impacts of policy interventions and infrastructure improvements. Initiatives such as implementing workplace mental health guidelines and enhancing university mental health facilities should be considered by researchers to fortify the support system within communities.

%emphasizing the responsibilities and broader community support needed. Efforts to create a non-stigmatized environment for self-disclosure through community capacity building have also been highlighted \cite{calear_sources_2022}.
%Technologies can broaden the scope of outreach and augment the cohesion of community members. 
%Community engagement to connect interpersonal-, technology-, and societal resources. 

\subsubsection{\textbf{Societal level}} Past investigations into societal-level resources have predominantly focused on professional help within the medical system, aligning with the formal help category in \textcite{rickwood_conceptual_2012}'s conceptual model.  

\textbf{Mitigate avoidance to seek professional help.} Our framework positions the societal level on the outermost layer, aligning with participants' natural order of help-seeking preferences \cite{gowen_online_2013}. The study revealed that 12 participants initially sought assistance from proximate resources in their daily lives, such as family and online information, before considering the societal level. This finding aligns with the pathway to psychiatric care model \cite{huxley_mental_1996}, shedding light on the avoidance process using the lived experiences of young adults. 
Barriers such as the limited availability of professionals, high costs, and challenges in finding suitable therapists contribute to this reluctance. 
%Young adults may access the societal level resources directly or through pathways that are mediated by other levels of resources. 

\textbf{Improve emergency services.} The National Crisis Line plays noticeable roles in \textit{\textit{care-giving}} and \textit{\textit{care-mediating}}. During emergencies when other levels of resources are inaccessible or unhelpful, the Crisis Line acts as the last line of defense \cite{pendse_can_2021}. The National Crisis Line serves as a crucial buffer for young adults facing emergencies, functioning as a significant channel for the \textit{care-mediating} mechanism. It directs callers to professional help, and in certain cases, enforces hospitalization. However, this approach has dual effects; for instance, one participant expressed fear of calling this number, highlighting concerns about the agency's response.

\textbf{Raise public awareness.} Efforts in \textit{\textit{care-outreaching}} at the societal level are escalating, especially with increased involvement in mental health awareness campaigns. Participants noted encountering relevant news and videos on social media platforms which raised their awareness \cite{liu_modeling_2023}. 
Notably, besides the medical system, none of our participants accessed other resources at this level, such as non-profit and government programs, indicating a scarcity of such support. The fact that two participants failed to locate these societal services underscores the urgency of drawing attention from public institutions to address this gap.

\subsection{Design implications for mental health technologies}
\label{sec: design}
Successful adoption and use of technologies require a holistic understanding of mental health technologies in people's social, cultural, and material environments \cite{gould_technology_2020}. We discussed how to augment the role of technologies via the three mechanisms in future design practices.
%This paper fills this gap by investigating young adults' use of resources to seek help for their mental health concerns in the context of their lived experiences and factors that influence the selection of the resources to shed light on designing 
%This study addresses this gap by examining how young adults utilize various resources in seeking help for mental health concerns, considering their lived experiences and influential factors. This research aims to contribute insights for designing effective digital MH tools with improved ecological validity.
%Studies by \cite{lattie_designing_2020} and \cite{le_exploring_2021} contextualized the needs of college students within their social ecosystems, co-designing technologies that seamlessly integrate into their daily lives to address anxiety. A systematic review and narrative synthesis by \cite{leamy_conceptual_2011} emphasized the roles of pathways and resources in the conceptual framework for personal recovery in mental health.
\subsubsection{Design for technology-mediated \textit{care-giving}}
Emotional support is one of the most important goals of using online communities \cite{metts_perceptions_2022}. 
%Future research is encouraged to explore how to balance personalization \cite{shalaby_peer_2020} and inclusiveness in online mental health communities. 

\textbf{Foster connection through shared experiences.}
Peer support plays a crucial role in enhancing user engagement with mobile mental health apps \cite{wong_postsecondary_2021, jonathan_smartphone-based_2021}. The informational value derived from others' narratives and shared experiences is more easily comprehended and accepted \cite{q_experiences_2021, liu_consumer_2023}. For instance, P15 expressed that reading about other people's journeys helped her navigate her own path to combat depression.
However, the current platforms do not fully cater to users' personalization needs. Some participants have a stringent definition of "similar experiences," seeking not only similarity but also a connection in life situations, such as socioeconomic status. For example, P05, who suffered from depression after a traumatizing gunshot experience, couldn't resonate with posts from wealthier individuals who had more resources and support, leading to a sense of exclusion. Research by \textcite{liang_embracing_2021} suggests that having a help provider with similar experiences of mental illness significantly facilitates mental health help-seeking.
To address these nuances, designers can explore segmenting subgroups on online communities based on stressors, creating experience-based subgroups, and involving peers in the early design stages \cite{g_peer_2022}.

%\textbf{Curate inclusive experience sharing on online communities.} 
%<Perceptions of Helpful and Unhelpful Responses to Disclosures of Suicidality in a Sample of Mobile App Users> 
%have highlighted the importance of resources aligned with the socio-economic situations and experiences of help providers to facilitate help-seeking \cite{gould_technology_2020}

\subsubsection{Design for technology-mediated \textit{care-mediating}}
The accessibility advantages of technologies hold promise in connecting and mediating various levels of resources. Technology-based services should establish connections back to family support, emphasizing intimate relationships that can play a semi-familial role \cite{palmier-claus_integrating_2013}. Nudging technology-mediated self-disclosure at the interpersonal level is crucial, offering a pathway to connect users with support systems.

\textbf{Develop technology-mediated self-disclosure.} 
The fear of disclosing mental distress hinders help-seeking and the establishment of support networks \cite{chung_medical_2020}. Three participants refrained from seeking help from interpersonal resources due to concerns about exposing vulnerability to them or considering mental distress "hard to talk about." Recognizing the potential of Large Language Models (LLMs) to accurately detect human emotions \cite{wang_reprompt_2023}, a promising avenue for future exploration involves designing mediators like LLM-empowered Chatbots \cite{jo_understanding_2023} to facilitate and encourage self-disclosure in intimate relationships. In professional settings, chatbots can potentially improve professional-patient communication and support self-disclosure to professionals such as therapists and psychiatrists. \textcite{koulouri_chatbots_2022} found young adults are likely to accept chatbots, and professionals confirmed that chatbots can be an assistant to raise awareness and mindfulness, elicit self-disclosure, and eliminate negative emotions \cite{lee_i_2020}.
%Sensors detecting the mood and trigger expression \cite{c_feasibility_2022}.

\textbf{Innovate technologies for enhancing intimate relationships.}  
%It is worth investigating the boundary between surveillance to prevent defensive behaviors \cite{ledbetter_parental_2010}.
The indirect mediating mechanism underscores the importance of intimacy and trust building through prolonged engagement \cite{shin_designing_2021}. Research by \textcite{dalsgaard_mediated_2006} identifies self-disclosure, trust, and commitment as key elements in constructing strong-tie intimacy. While family and friends serve as a natural choice for individuals with strong interpersonal relationships, participants lacking intimacy and trust face challenges and have to win for understanding. Technologies play a crucial role in supporting the building of a robust support system both before and during the help-seeking process, cultivating strong social support to overcome resistance and encourage seeking professional help \cite{gould_technology_2020}. 
In particular, gaming has been identified as an effective means of fostering connection and providing an avenue for emotional self-disclosure \cite{musick_gaming_2021}. Four male participants in our study leaned on gaming to stay connected with friends and exchange thoughts. Technologies can facilitate the creation of a safe and private communication channel \cite{wong-villacres_technology-mediated_2011}.

\subsubsection{Design for technology-mediated \textit{care-outreaching}}
Currently, technologies play less important roles in \textit{care-outreaching}. 

\textbf{Reach out to young adults with low-technologies.} Only two participants in our study attempted to use Mhealth apps. This aligns with prior findings indicating a limited uptake rate and efficacy of real-world deployment despite the increasing prevalence of self-management apps \cite{burke_qualitative_2022}. \textcite{gitlow_how_2019} distinguishes Mhealth technologies into high-tech and low-tech categories, with low technology encompassing simple, non-complex tools such as text messages and notifications. Users perceive low technologies as more acceptable, efficient, flexible, and easily integrated into daily life, facilitating help-seeking for mental distress \cite{gould_technology_2020}. Four participants expressed that emails, posters, and social media messages from the university effectively reminded and encouraged them to utilize university services. Additionally, three participants became aware of the National Crisis Line through top results on platforms like YouTube and Google. This highlights the significance of implementing similar nudges on other widely used platforms to enhance awareness and accessibility to essential resources.
Notably, low technology, especially in urgent contexts like suicidal attempts, plays a crucial role by providing consistently available and timely responses \cite{palmier-claus_integrating_2013}. It offers more accessible and equitable services, particularly benefiting low-resource populations \cite{r_country_2021}. 

%For instance, \textcite{le_exploring_2021} discovered that smartphones can be seamlessly incorporated into self-regulation practices. Rather than creating new tools and apps, the emphasis should be on implementing and embedding everyday, low-tech solutions into young adults' sociotechnological ecosystems \cite{chen_information_2015}. 

\textbf{Enhance proactive and long-term intervention tools in professional treatment.}
In our study, eight participants' help-seeking was initiated by active intervention from others such as interpersonal and community resources. 
Technologies capable of monitoring and identifying early signs of depression and anxiety can proactively engage young adults living independently, issuing warnings and prompting help-seeking when needed. Future studies should prioritize user involvement in the early stages of design \cite{alqahtani_co-designing_2021} to ensure effective adoption \cite{danieli_conversational_2021, zhang_design_2023}. Integrating cutting-edge technologies, such as asynchronous remote support \cite{bhattacharya_designing_2021} and multimodal communication \cite{h_barriers_2021}, holds the potential to enhance user engagement in professional treatment contexts.

\section{Conclusion}
This study conducted 18 in-depth interviews exploring the lived experiences of mental health help-seeking. We proposed the Socio-technical Ecosystem Framework of mental health resources, grounded in data analysis and the Social Ecological Theory. This framework delineates a four-level resource ecosystem, details three mechanisms of resource support, classifies two types of support system, and outlines diverse pathways to care within this socio-technical ecosystem. Through the framework, we identified barriers at each level and overarching challenges in establishing a support system. In the discussion, we underscore the novel contributions of this framework compared to previous models, outline a structured research agenda, and offer implications for the future development of mental health technologies

\bibliographystyle{ACM-Reference-Format}
\bibliography{references}

\appendix
\end{document}